\documentclass{aa}
\usepackage{txfonts}
\usepackage{graphicx}

\bibpunct{(}{)}{;}{a}{}{,}

\begin{document}

\title{Evidence of amplitude modulation due to resonant mode coupling \\ in the $\delta$ Scuti star KIC~5892969}
\subtitle{A particular or a general case?}

\author{{Barcel\'o Forteza}, S.\inst{\ref{ins1},\ref{ins1.1}} \and Michel, E.\inst{\ref{ins2}} \and {Roca Cort\'es}, T.\inst{\ref{ins1},\ref{ins1.1}} \and Garc\'ia, R.~A.\inst{\ref{ins3}} }

\institute{Instituto de Astrof\'isica de Canarias, 38200 La Laguna, Tenerife, Spain; e-mail: sebastia@iac.es \label{ins1}
\and
Departamento de Astrof\'isica, Universidad de La Laguna, 38206 La Laguna, Tenerife, Spain \label{ins1.1}
\and
LESIA, Observatoire de Paris, CNRS (UMR 8109), Université Pierre et Marie Curie, Université Denis Diderot, Pl. J. Janssen, 92195 Meudon\label{ins2}
\and
Laboratoire AIM, CEA/DSM – CNRS – Univ. Paris Diderot – IRFU/SAp, Centre de Saclay, 91191 Gif-sur-Yvette Cedex, France\label{ins3}
}

\date{Received 12 December 2014; Accepted 28 May 2015}

\abstract{A study of the star KIC~5892969 observed by the \textit{Kepler} satellite is presented. Its three highest amplitude modes present a strong amplitude modulation. The aim of this work is to investigate amplitude variations in this star and their possible cause.\\ Using the 4 years-long observations available, we obtained the frequency content of the full light curve. Then, we studied the amplitude and phase variations with time using shorter time stamps.\\ The results obtained are compared with the predicted ones for resonant mode coupling of an unstable mode with lower frequency stable modes. Our conclusion is that resonant mode coupling is consistent as an amplitude limitation mechanism in several modes of KIC~5892969 and we discuss to which extent it might play an important role for other $\delta$ Scuti stars.}


\maketitle

\section{Introduction}
\label{s:intro}

\paragraph{}While the instability of the modes in $\delta$ Scuti stars is well understood as mainly due to $\kappa$-mechanism of the HeII ionization zone \citep{Pamyatnykh1999}, the mechanism of amplitude limitation is a long standing open question. This mechanism has been studied theoretically by several authors \citep[e.g.][]{Vandakurov1979, Dziembowski1982}, however, observations that are long and precise enough to test these possible causes have been very limited so far \citep[see][for a detailed review]{DSRS}.

\paragraph{}Using ground-based observations, several $\delta$ Scuti stars have long been studied: 4 Canum Venaticorum \citep[4CVn;][]{Breger1990} has been followed by the Delta Scuti Network for decades allowing the observation of amplitude modulation in several peaks as well as a change in phase in one of them. XX Pyx \citep{Handler2000} has also been observed with the Delta Scuti Network, and it presents cyclical amplitude and phase variations in several oscillation modes. Mode coupling was in both cases considered the explanation of these variations \citep[e.g.][]{Nowakowski2005}, but this could not be ascertained.

\paragraph{}Photometric observations by the recent space missions, \textit{CoRoT} \citep{Baglin2006} and \textit{Kepler} \citep{Borucki2010}, represent a good opportunity for the study of the amplitude limitation mechanism: the long duration of the observing runs (up to four years in the case of the \textit{Kepler} satellite) and their high duty cycle allowed us to study the amplitude modulation in the characteristic modes of the stars with better precision than before.

\paragraph{}One interesting case is the $\delta$ Scuti star KIC~8054146 observed for four years with \textit{Kepler}. \citet{Breger2014} found several low-amplitude modes showing amplitude and phase variations. They show that these variations follow a non-linear relation attesting their coupling. In addition, they were able to differentiate which modes are parent or child. \citet{Bowman2014} studied the $\delta$ Scuti star KIC~7106205 that presents one mode showing an amplitude decrease of one order of magnitude and a significant phase change. The authors stress that mode coupling is one of the possible reasons for that damping.

\paragraph{}Star KIC~5892969 is another interesting case of amplitude modulation that we study in this paper. In Section \ref{s:tomberi} we describe the main characteristics of this star and the way in which its oscillations are analysed. Its amplitude and phase variations with time are studied in Section \ref{s:pieces}. In Section \ref{s:modulation} we show that resonant mode coupling can explain the observed amplitude modulation in the oscillating modes. In the Section \ref{s:discussion} we discuss our results. In the last section we present our conclusions and compare them with other studies.

\section{Analysis of KIC~5892969 light curve}
\label{s:tomberi}

\paragraph{}KIC~5892969 is a faint $\delta$ Scuti star whose characteristics are detailed in Table \ref{t:stelchar}. Delta Scuti type stars are classical pulsators excited by $\kappa$-mechanism, located on or just off the main sequence, with masses between 1.5 and 2.5 $M_{\odot}$ \citep{Breger2000}. They show fast rotation rates as it is common in stars with these masses or higher \citep{Royer2007}.

\begin{table}
\caption{KIC~5892969 stellar parameters \citep{Huber2014}}
\label{t:stelchar}
\centering
\begin{tabular}{c c c c c}
\hline\hline
\multicolumn{5}{c}{KIC~5892969}\\
\hline
$Kp$&& 12.445&&-\\
$T_{\mathrm{eff}}$&$(\mathrm{K})$& 7560& $\pm$& 240/300\\
$\mathrm{log} ~\textit{g}$& $(\mathrm{cm/s^{2}})$& 3.76& $\pm$& 0.26/0.12\\
$[\mathrm{Fe}/\mathrm{H}]$& $(\mathrm{dex})$& -0.10& $\pm$& 0.24/0.36\\
$R$& $(\mathrm{R_{\odot}})$& 3.1& $\pm$& 0.6/1.0\\
$M$& $(\mathrm{M_{\odot}})$& 2.0& $\pm$& 0.2/0.4\\
\hline
\end{tabular}
\end{table}

\begin{figure}
\resizebox{\hsize}{!}{\includegraphics{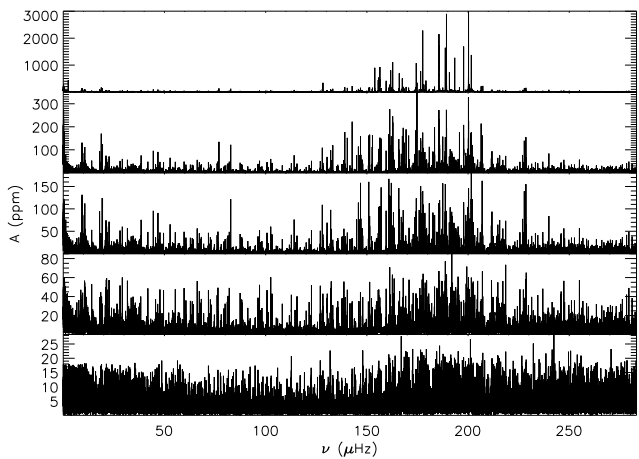}}
\caption{From top to bottom: amplitude spectra of the four-year KIC~5892969 original light curve and also those after extracting 50, 100, 250, and 1313 peaks. Notice the two order of magnitude change of scale between top and bottom graphs.}
\label{f:spectra}
\end{figure}

\paragraph{}The typical oscillation range of $\delta$ Scuti stars is between 58 and 580 $\mu$Hz although some oscillations have been found with frequencies up to 970 $\mu$Hz \citep{Zwintz2013}. Their power spectrum shows a complex structure with dominant peaks of moderate amplitudes and hundreds of lower amplitude peaks \citep[e.g.][]{Poretti2009}. This behaviour is also clearly observed in KIC~5892969 spectra (see Fig. \ref{f:spectra}) when our analysis is applied.

\subsection{\textbf{Methodology of the analysis}}
\label{s:method}

\paragraph{}KIC~5892969 has been observed during four years in long cadence (LC). That means a sampling time of 29.5 minutes in the satellite time frame. In fact, this sampling time shows a tiny sinusoidal variation in the barycentric time frame \citep{Garcia2014} that is of importance in Section \ref{s:alias}.

\paragraph{}We use simple aperture photometry (SAP) time series \citep{Thompson2013} corrected for outliers, jumps, and drifts, following the methods described in \citet{Garcia2011}. We study the full four-year light curve to obtain a precise value of the parameters for each peak with a frequency resolution of $\approx$8nHz. The analysis of each oscillation is done with a technique adapted from iterative sine wave fitting \citep[ISWF;][]{Ponman1981} considering that the light curve is a sum of sinusoidal terms, i.e.
\begin{equation}
F\;=\;\sum_{i}\; A_{i} \sin\left(2\pi\nu_{i} t + \phi_{i} \right)\, ,
\label{e:flux}
\end{equation}
where $A_{i}$, $\nu_{i}$ and $\phi_{i}$ are the amplitude, the frequency, and the phase of each peak respectively.

\paragraph{}In the first stage, we look at the highest amplitude peak. A guess of its parameters is obtained making a fast Fourier transform (FFT) of the time series where the gaps are filled with a linear interpolation. A fitting in the time domain is done exploring an oversampled grid of frequencies around the guess frequency. We only take the original observed points into account. The final values of the parameters of the peak are those associated with the highest amplitude. Then, we extract this sine wave from the light curve. 

\paragraph{}This process is repeated with the next highest peak until one of these three conditions is accomplished:\\
\textbf{1)}The computed root mean square of the residual signal increases instead of decreases.\\
\textbf{2)}The local signal-to-noise ratio (SNR) is lower than 3.0. The SNR is calculated comparing the amplitude of the extracted peak with the mean of the amplitudes in a range of 60 $\mu$Hz centred around this peak.\\
\textbf{3)}The fitting of the amplitude of the peak has a significant error ($\geq$20\%).

\paragraph{}In the second stage, we restart the process from the beginning. However, this time we obtain guess parameters from the FFT of the light curve where gaps are interpolated with a non-linear interpolation taking the parameters obtained in stage 1 into account. This stage, with those improved guesses, reduces the influence of artefacts due to the spectral window and allow us to go further in the analysis of lower amplitude peaks.

\paragraph{}The third stage of the method consists in repeating the process again. We obtain the guess parameters from the FFT of the light curve where gaps are interpolated according to parameters obtained at stage 2. This time the fit is done considering the full light curve where gaps have been interpolated. It improves the analysis further and take more interactions between peaks into account.

\paragraph{}We tested this method with an artificial light curve based on the parameters of the peaks of a well-known $\delta$ Scuti star (see Appendix \ref{ap:ALC}). This method is found to be interesting in various respects. It brings very accurate results in terms of frequencies, amplitudes, and phases while dealing with a large number of peaks; also, it is reasonably fast. In our understanding this comes from the fact that each optimization is dealing with a limited number of parameters, while the mutual influence of peaks is taken into account in the successive stages of the process. We believe that the efficiency of this method is also due to the very good observational window of space missions and high SNR.

\subsection{Frequency content of the full light curve}
\label{s:wholeres}

\paragraph{}When this method is applied to KIC~5892969, we obtain 1313 peaks higher than 20 parts per million (ppm) with a SNR greater or equal to 5 (noise has been calculated as we mention in Section \ref{s:method}). These peaks carry 98.8\% of the full signal. Fifteen peaks with amplitude higher than 1000 ppm are shown in Table \ref{t:peaks}. The analysis also finds around 1100 peaks under an amplitude of 80 ppm. These peaks carry only 1.7\% of the energy of the signal. As we mentioned at the beginning of Section \ref{s:tomberi}, this characteristic structure (see Fig. \ref{f:spectra}) has also been observed in other $\delta$ Scuti stars. The cause of this detected high density of peaks is unclear and still remains under debate \citep[e.g.][]{Balona2011,Mantegazza2012}.

\begin{table}
\caption{Peaks with amplitudes higher than 1000 parts per million, which have been identified in the spectrum of the whole light curve. The column Energy shows the amount of energy of the observed signal carried by the wave.}
\label{t:peaks}
\centering
\begin{tabular}{c c c c c c c}
\hline\hline
Term&Frequency &Amplitude&Energy&Split peaks\tablefootmark{\textbf{a}}\\
i&($\mu$Hz)&(ppm)&(\%)&\\
\hline
1&       200.22576&       3207& 12.0&\\
2&       185.61812&       3297& 12.7&\\
3&       189.32948&       2951& 10.2&\\
4&       177.59241&       2471& 7.1&\\  
5&       189.31940&       2299& 6.2&$f_{3}$-$s_{RMC,3}$\\
6&       200.21537&       2379& 6.6&$f_{1}$-$s_{RMC,1}$\\
7&       185.62741&       1779& 3.7&$f_{2}$+$s_{RMC,2}$\\
8&       197.79423&       1983& 4.6&\\
9&       188.73938&       1649& 3.2&\\
10&      201.59579&       1647& 3.2&\\
11&      193.39818&       1327& 2.1&\\
12&      153.91426&       1382& 2.2&\\
13&      174.39755&       1108& 1.4&\\
14&      162.75762&       1108& 0.8&\\
17&      156.43953&       1172& 1.6&\\
\hline
\end{tabular}
\tablefoot{\tablefoottext{\textbf{a}}{The marked peaks are the first order split peaks (see Section \textbf{\ref{s:splitting}}) of their respective modes. The difference between them and the main peak ($s_{RMC}$) is shown in Table \ref{t:modpar}.}}
\end{table}

\section{Amplitude and phase variations with time}
\label{s:pieces}

\paragraph{}Thanks to the four-year duration of the observations, it is possible to study shorter subseries to unveil any variations of the parameters for each mode with time. Using the frequencies obtained in Section \ref{s:wholeres}, we fit the amplitude and phase of each peak in 100 days subseries shifted every 10 days. The duration of each piece and the shift in time is chosen to reach a compromise between time and frequency resolution. Tests on artificial signal show a maximum drift of 0.01 radians when we refer the phase to the initial time value of the observation. Moreover, we detect no significant errors in amplitude.

\begin{figure}
\resizebox{\hsize}{!}{\includegraphics{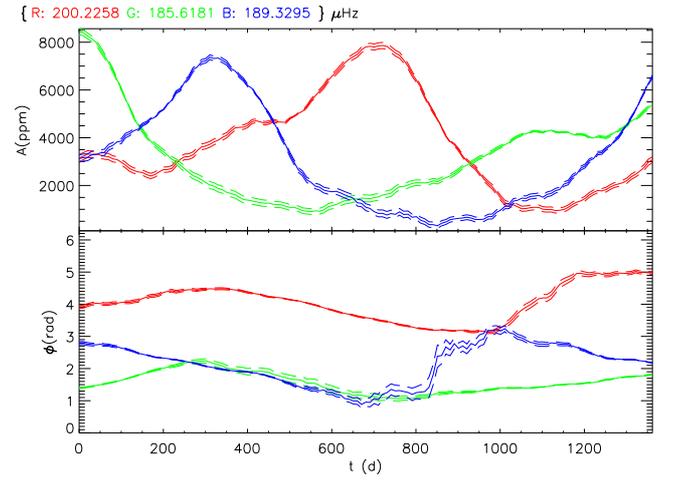}}
\caption{Amplitude (top) and phase (bottom) variations with time in days from the beginning of the observations. Red, green, and blue denote modes $f_{1}$, $f_{2}$, and $f_{3}$, respectively. Dashed curves are drawn at $\pm 1 \sigma$ error bars.}
\label{f:mod3}
\end{figure}

\paragraph{}This analysis reveals a striking pattern in the three highest amplitude modes: they show a modulation of about one order of magnitude in their amplitude, with a slower increase than decrease (see Fig. \ref{f:mod3}). This faster amplitude decrease is followed by a sharp change in phase. In addition, these three modes show the same shaped variation but shifted in time.

\begin{figure}
\resizebox{\hsize}{!}{\includegraphics{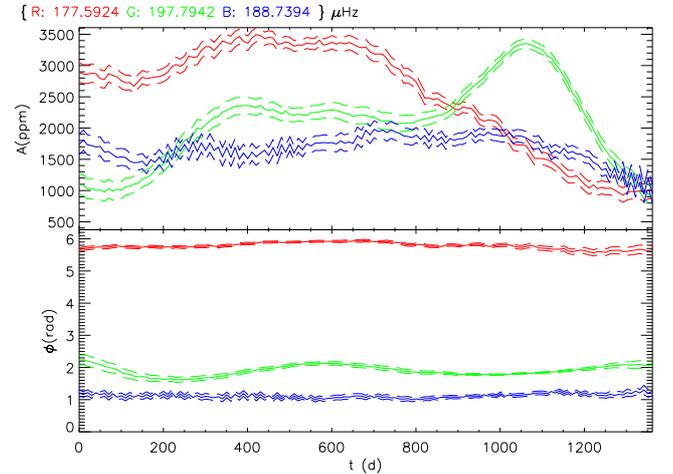}}
\caption{Same as Fig. \ref{f:mod3} for modes $f_{4}$, $f_{8}$, and $f_{9}$ (red, green, and blue, respectively) showing small amplitude change.}
\label{f:mod678}
\end{figure}

\begin{figure}
\resizebox{\hsize}{!}{\includegraphics{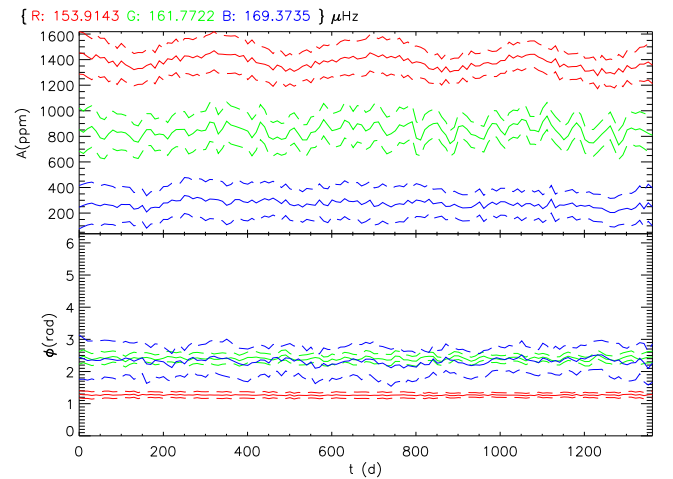}}
\caption{Same as Fig. \ref{f:mod3} for modes $f_{12}$, $f_{15}$, and $f_{55}$ (red, green, and blue, respectively) showing slight amplitude change.}
\label{f:nomod}
\end{figure}

\begin{figure}
\resizebox{\hsize}{!}{\includegraphics{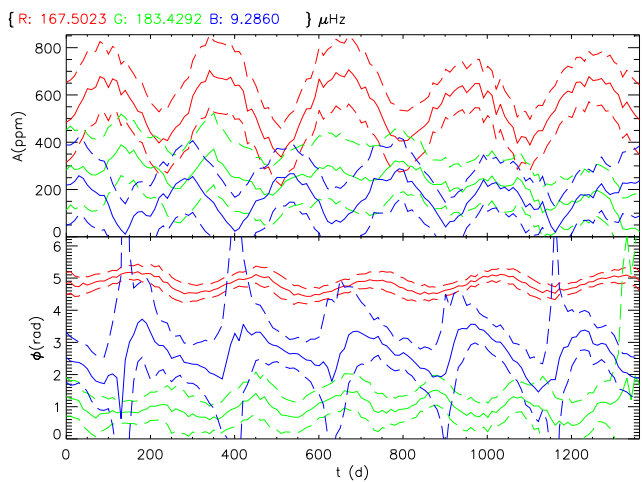}}
\caption{Same as Fig. \ref{f:mod3} for modes $f_{31}$, $f_{78}$, and $f_{103}$ (red, green, and blue, respectively). The amplitude variation of $f_{99}$ is of 1 order of magnitude approximately. The period of these modulations is approximately a submultiple of the one of peaks $f_{1}$, $f_{2}$, and $f_{3}$.}
\label{f:modshort}
\end{figure}

\paragraph{}Several other analysed peaks show smaller variations or no detectable modulation at all (see Figures \ref{f:mod678} and \ref{f:nomod}), while many others show variations in shorter timescales (see Fig. \ref{f:modshort}).

\section{Investigating the modulation mechanism}
\label{s:modulation}

\paragraph{}The most appealing interpretation for these amplitude and phase modulations is the resonant mode coupling as it is described in \citet{Moskalik1985}. As we already mentioned in Section \ref{s:intro}, resonant mode coupling is one of the possible causes considered for amplitude limitation in $\delta$ Scuti stars. We summarize its main lines in Section \ref{s:coupling}. Then, we discuss in Section \ref{s:ocauses} other possible causes for amplitude and phase variation and explain why they are discarded here.

\subsection{Resonant mode coupling}
\label{s:coupling}

\paragraph{}There are several different cases of resonant interaction between modes depending on their characteristics. In \citet{Moskalik1985}, the case of one unstable mode coupled with two lower frequency stable modes is analysed. Saturation of the driving mechanisms and non-adiabaticity are neglected. This study considers three modes satisfying the near-resonance condition:
\begin{equation}
\nu_{0} = \nu_{1}+\nu_{2}-\delta; \; |\delta| \ll \nu_{i}\, ,
\label{e:nrcond}
\end{equation}
where $\nu_{0}$ is the linear frequency of the "child" mode and $\nu_{1,2}$ the "parent" modes. These two parent modes are assumed to be stable and have the same damping rate.

\paragraph{}The behaviour of the amplitude and phase of the coupled modes is cyclic with a certain periodicity, $P$. The child mode shows slowly increasing amplitude and a quiet phase. After the amplitude has reached the maximum, it decreases rapidly and it is accompanied by a violent phase change \citep[see Fig. 2 in][]{Moskalik1985} as we see for modes $f_{1}$, $f_{2}$, and $f_{3}$ in Fig. \ref{f:mod3}. The characteristic times of growing ($t_{g}$) and shrinking ($t_{d}$) depends on the driving rate of the unstable mode ($\gamma_{0}$) and the driving rate of the stable modes ($\gamma$).

\paragraph{}The depth of the amplitude modulation ($A_{max}/A_{min}$) is determined by the ratio of the frequency mismatch ($\delta$) and the driving rates. A higher rate produces a lower decay \citep[see Fig. 4 in][]{Moskalik1985}. The change of phase during the decrease of amplitude ($\Delta\phi$) of the mode can be greater than $\pi /2$ in absolute value and it has the same sign as the frequency mismatch. The variation in phase depends on the frequency mismatch in the following way:
\begin{equation}
\delta = \left< \dfrac{d\phi_{0}}{dt}\right>_{P} - \left<\dfrac{d\phi_{1}}{dt}\right>_{P} - \left<\dfrac{d\phi_{2}}{dt}\right>_{P}\, .
\label{e:totalphase}
\end{equation}
This relation points that the phase could not be completely periodic \citep{Moskalik1985}.

\paragraph{}The results found in our analysis are consistent with those obtained by \citet{Moskalik1985}, at least for the three highest modes. The observed amplitude variations are approximately of one order of magnitude and the phase changes are greater than $\pi /2$ in two of the modes (see Table \ref{t:modpar}).

\begin{table*}
\caption{Characteristic parameters of the modulated peaks compared with the typical values for modes of $\delta$ Scuti stars \citep{Moskalik1985}.}
\label{t:modpar}
\centering
\begin{tabular}{c c c c c c}
\hline\hline
&&$\delta$ Scuti&$f_{1}$&$f_{2}$&$f_{3}$\\
\hline
$\nu_{k=0,i}^{\prime}$    &          &               &    200.22576    &     185.61812    &    189.32948   \\
$\nu_{i}$                 &          &               &    200.22096    &     185.62042    &    189.32536   \\
$A_{max}/A_{min}$         &          &               &  8.3 $\pm$ 0.9  &   9.3 $\pm$ 1.3  & 23.4 $\pm$ 8.2 \\
$\Delta\phi$              & (rad)    &               & 2.08 $\pm$ 0.19 & -1.24 $\pm$ 0.05 & 2.21 $\pm$ 0.17\\
$\delta\nu / \nu$         & $10^{-5}$& $\lesssim$100 & 13.4 $\pm$ 0.4  &  -8.5 $\pm$ 0.1  & 24.1 $\pm$ 1.5 \\
$t_{d}$                   & (days)   & [15,700]      &  411 $\pm$ 20   &   551 $\pm$ 20   &  541 $\pm$ 20  \\
$P$                       & (days)   & $\sim$1500    &  961 $\pm$ 20   &  1362 $\pm$ 20   & 1362 $\pm$ 20  \\
$\gamma_{0}$              & (nHz)    & $\sim$8       & 12.0 $\pm$ 0.3  &   8.5 $\pm$ 0.1  &  8.5 $\pm$ 0.1 \\
$s_{RMC}$                 & (nHz)    &               & 10.4 $\pm$ 0.1  &   9.3 $\pm$ 0.2  & 10.1 $\pm$ 0.1 \\
$max(A_{|k|=1})/A_{k=0}$  &          &               & 0.74 $\pm$ 0.01 &  0.54 $\pm$ 0.01 & 0.78 $\pm$ 0.01\\
$h_{-1}/h_{+1}$           &          &               & 4.33 $\pm$ 0.10 &  0.37 $\pm$ 0.12 & 2.78 $\pm$ 0.05\\
\hline
\end{tabular}
\end{table*}

\paragraph{}Moreover, the driving rate can be roughly estimated from the period of the amplitude modulation. The typical driving rate for $\delta$ Scuti stars is of the same order as those calculated here from the amplitude modulation (see Table \ref{t:modpar}). In fact, Fig. \ref{f:mod3} suggests that the observations encompass approximately one period ($P$). For the modes $f_{2}$ and $f_{3}$, however, we mention values of $P$ as lower limits.

\paragraph{}The amplitude and phase variations of the modes $f_{31}$, $f_{78}$, and $f_{103}$ in Fig. \ref{f:modshort} show a cyclic behaviour, which can also be explained by mode coupling. They can be modes with higher driving rates and shorter period of modulation or they can also be modes with the same period but showing several maxima in one cycle as described in \citet{Wersinger1980}. It seems difficult to exclude one option instead of the other; however, the second option is assumed to be rare in real stars \citep{Moskalik1985}.

\paragraph{}In addition, the variation in phase produces a change in frequency during the fast decrease of the amplitude \citep{Moskalik1985}, i.e.
\begin{equation}
\delta\nu = (2\pi)^{-1}\cdot \left.\dfrac{d\phi}{dt}\right\vert_{t}\, .
\label{e:nosplit}
\end{equation}

\paragraph{}The relative change in frequency ($\frac{\delta\nu}{\nu}$) that we measured is compatible with that \citet{Moskalik1985} estimates for $\delta$ Scuti stars (see Table \ref{t:modpar}) and its sign is in agreement with the sign of the phase change \textbf{($\Delta\phi$)}.

\subsubsection{Multiplet of the resonant coupling}
\label{s:splitting}

\paragraph{}Following \citet{Moskalik1985}, the amplitude and phase variations of a given peak can also be observed in the spectrum of the whole light curve. There, each peak appears with several nearby components characterizing the variations with time. The splitting of the resonant mode coupling ($s_{RMC}$) can be defined as the frequency separation between the components of the observed multiplet
\begin{equation}
s_{RMC} \equiv (\mathbf{\nu_{k\neq 0}^{\prime}}-\mathbf{\nu_{k=0}^{\prime}})/k\, .
\label{e:split1}
\end{equation}
We estimated the value of this splitting considering the first order peaks ($|k|=1$) as the higher peaks (see Table \ref{t:modpar}).

\begin{figure}
\resizebox{\hsize}{!}{\includegraphics{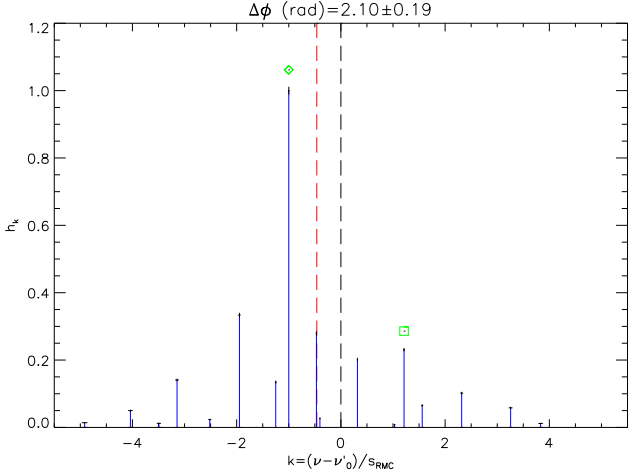}}
\caption{Peaks near the unstable mode $f_{1}$. The abscissa represents the frequency difference from the central peak (black dashed line) in units of the splitting ($s_{RMC}$). The amplitudes have been normalized to the highest first order split peak. Green diamond and square are pointing to the maximum and the minimum first order split peaks respectively. The linear frequency of the mode (red dashed line) is lower than the central peak (see text).}
\label{f:som1}
\end{figure}

\begin{figure}
\resizebox{\hsize}{!}{\includegraphics{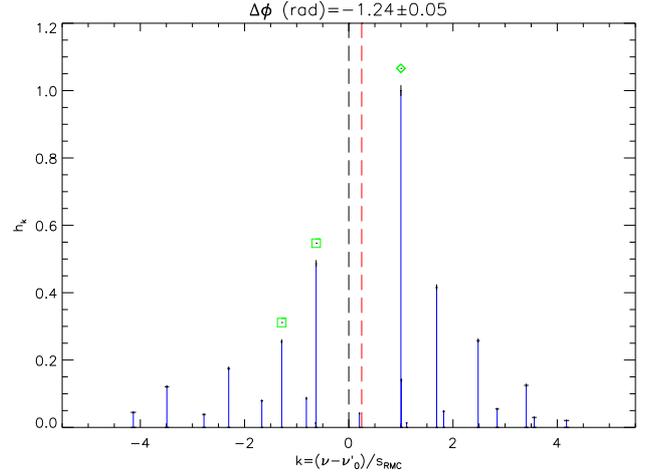}}
\caption{Same as Fig. \ref{f:som1} for $f_{2}$. The mode presents opposite asymmetry and sign in $\Delta\phi$ than in the mode $f_{1}$.}
\label{f:som2}
\end{figure}

\begin{figure}
\resizebox{\hsize}{!}{\includegraphics{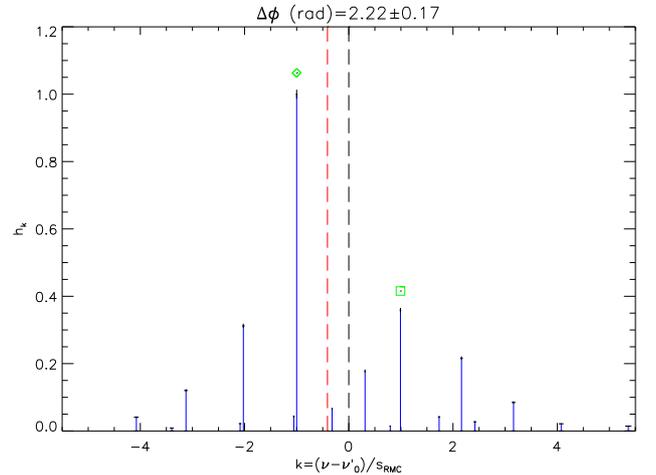}}
\caption{Same as Fig. \ref{f:som1} and \ref{f:som2} for the mode $f_{3}$. We can observe the same asymmetry as in $f_{1}$ and opposite from $f_{2}$. This is in agreement with the sign in $\Delta\phi$.}
\label{f:som3}
\end{figure}

\paragraph{}These multiplets show right- or left-handed amplitude asymmetry depending on the sign of the frequency mismatch (see Eq. \ref{e:totalphase}) and, therefore, on the sign of the phase change (see Figures \ref{f:som1}, \ref{f:som2}, and \ref{f:som3}). This asymmetry can be characterized by the ratio of amplitudes of the multiplet with the highest first order peak, i.e.
\begin{equation}
h_{k}=A_{k}/max(A_{\pm 1})\, .
\label{e:aratios}
\end{equation}

\paragraph{}Particularly, $h_{-1}/h_{+1}$ is the best indicator: it shows a value lower than one for right-handed asymmetry and a value higher than one for left-handed asymmetry. As expected, Table \ref{t:modpar} shows that the sign of phase change is related to this asymmetry.

\paragraph{}The linear frequency of the mode ($\nu_{i}$) is related to the observed one ($\nu_{k=0,i}^{\prime}$) through the non-periodicities \citep[see][]{Moskalik1985}:
\begin{equation}
 \nu_{i}= \nu_{k=0,i}^{\prime}-\left< \dfrac{d\phi_{i}}{dt}\right>_{P}\, .
\label{e:lf}
\end{equation}
This parameter is another measurement of the asymmetry of the mode. As shown in Figures \ref{f:som1} to \ref{f:som3}, the linear frequency is lower or higher than the observed frequency depending on the sign of phase change ($\Delta\phi$).

\subsubsection{Parent-child relations}
\label{s:parent}

\paragraph{}According to \citet{Moskalik1985}, the parent modes are expected to have very small amplitudes. When the amplitude of the child mode reaches a certain value the parent amplitudes start to grow rapidly and the amplitude of their child mode decreases. After that, the parent modes amplitude decrease \citep[see Fig. 2 in][]{Moskalik1985}. He also points out that the parent modes of real stars might have different driving rates ($\gamma_{1}\neq\gamma_{2}$). However, his study shows that there is no expectation of inducing any significant change in the behaviour: the amplitudes of the parents are just different because of the stronger damping of one respect to the other.

\paragraph{}Although the uncertainties in the parameters of these stable modes and the lower amplitude of their split peaks makes it difficult to detect, it would be possible to find parent and child mode relations. Looking for three modes that follow Eq. (\ref{e:nrcond}) and (\ref{e:totalphase}) at the same time, which taking into account Eq. (\ref{e:lf}) gives
\begin{equation}
\nu_{0}^{\prime} = \nu_{1}^{\prime} + \nu_{2}^{\prime} \, .
\label{e:tp2}
\end{equation}
Therefore, to be in a parent-child relation, the modes have to fulfill two conditions:\\
\textbf{1)}The observed frequencies of these peaks exactly satisfy Eq. \ref{e:nrcond}. It would mean that $\mathbf{\delta^{\prime}=\nu_{1}^{\prime}+\nu_{2}^{\prime}-\nu_{0}^{\prime}}$ has to be 0 (within errors).\\
\textbf{2)}The frequency mismatch of these modes ($\delta$) has the same sign of phase change ($\mathbf{\Delta\phi}$).

\paragraph{}Our analysis allowed us to find four possible parent-child relations for mode $f_{1}$, two for mode $f_{2}$ and one for mode $f_{3}$ (see Table \ref{t:ppchr}). As could be seen, their frequency mismatch is in agreement with the change of phase of the child mode.  

\begin{table}
\caption{Parent-child relations for the three highest amplitude modes (see text).}
\label{t:ppchr}
\centering
\begin{tabular}{c c c c c c c}
\hline\hline
$f_{i}$&$\Delta\phi$&Possible&$\delta^{\prime}$&$\delta$\\
&(rad)&parents&(nHz)&(nHz)&\\
\hline
1&  2.08 $\pm$ 0.19 & $f_{14}+f_{680}$  &  0.16 &  2.2 $\pm$ 2.4\\
 &                  & $f_{34}+f_{309}$  &  0.32 &  4.0 $\pm$ 1.5\\
 &                  & $f_{168}+f_{112}$ &  0.16 &  2.2 $\pm$ 5.1\\
 &                  & $f_{83}+f_{1061}$ & -0.23 &  1.8 $\pm$ 3.4\\\hline
2& -1.24 $\pm$ 0.05 & $f_{12}+f_{266}$  & -0.16 & -2.5 $\pm$ 0.6\\
 &                  & $f_{17}+f_{262}$  &  0.23 & -1.3 $\pm$ 0.5\\\hline
3&  2.21 $\pm$ 0.17 & $f_{2}+f_{953}$   &  0.08 &  2.5 $\pm$ 3.9\\
\hline
\end{tabular}
\end{table}

\begin{figure}
\resizebox{\hsize}{!}{\includegraphics{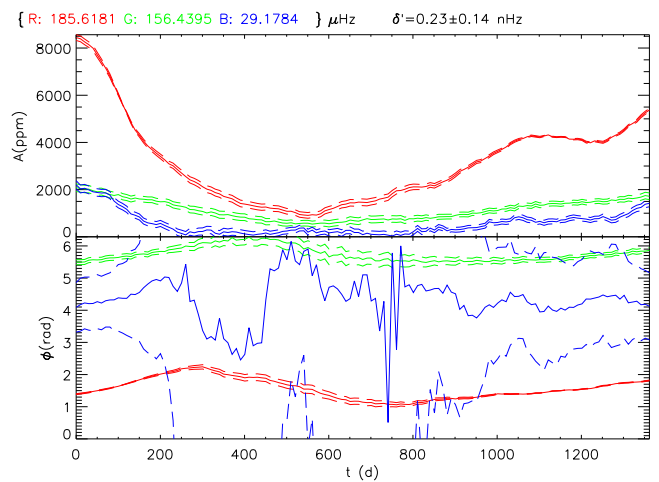}}
\caption{Amplitude and phase variations with time. Red, green and blue for modes $f_{2}$, $f_{17}$, and $f_{262}$, respectively. The amplitude of $f_{262}$ has been increased by one order of magnitude to properly observe its variations. The parent modes are $f_{17}$, and $f_{262}$, while $f_{2}$ is the child mode (see text).}
\label{f:cm1}
\end{figure}

\paragraph{}One example of possible coupled modes is shown in Fig. \ref{f:cm1}. In this case, the supposed parent modes ($f_{17}$, and $f_{262}$) would have different dampings, as shown by the fact that one of them ($f_{262}$) reaches its minima faster than the other ($f_{17}$). However, they reach their maxima at the same time, when the child mode (\textbf{$f_{2}$}) starts to decrease as expected. We observe a difference in amplitude of the two parent modes which is coherent with the fact that they might have different driving rates. In this case, the phases of these modes vary at the same time. This is yet a third condition for a parent-child relation.

\paragraph{}Beyond the three-modes case, it is possible that more than three modes are coupled. \citet{Nowakowski2005} points to a statistical equilibrium of the modes when many pairs of g modes exchange energy with the driving rate of an acoustic one. The number of parent peaks involved in that case has to be at least four. Furthermore, if their coupling coefficients are close enough, they might get synchronized and interact like a single pair. This would also be compatible with the fact that our analysis is able to find several possible parent-child relations.

\subsection{Rejecting other possible causes for amplitude modulation}
\label{s:ocauses}

\paragraph{}Resonant mode coupling is not, a priori, the only possible cause of the variation of the measured parameters of the modes. Here we discuss some other possible causes that have been ruled out as causes of the observed variations for modes $f_{1}$, $f_{2}$ and $f_{3}$.

\subsubsection{Binarity}
\label{s:binarity}

\paragraph{}The presence of a companion body (second star or exoplanet) might induce phase shifts in the modes due to the Doppler effect. This appears as a frequency splitting in the power spectrum of the whole light curve \citep{ShK2012}.

\paragraph{}However, we discarded this effect as the main cause of the observed variations in KIC~5892969 because it is not expected to produce amplitude modulation and it is supposed that the phase changes are synchronized among all the different modes \citep{Murphy2014}, which is not the case here (see Fig. \ref{f:mod3} to \ref{f:nomod}).

\paragraph{}In this respect, the recent discovery that 4CVn is an eccentric binary system \citep{Schmid2014} cannot explain the amplitude variations observed by \citet{Breger1990}, also suggesting that its mode variations might be due to mode coupling \citep{Breger2000a}.

\subsubsection{Interferences with superNyquist peaks}
\label{s:alias}

\paragraph{}Variations of amplitude and phase with time could be due to interferences between real peaks close to each other as we see in Section \ref{s:splitting}. However, this variation can also be due to an artefact created by a peak located above the Nyquist frequency (superNyquist peak) close to a real one.

\paragraph{}LC \textit{Kepler} data has a Nyquist frequency \textbf{($\nu_{Ny}$)} of 283 $\mu$Hz approximately. The higher limit of the typical range of frequencies in $\delta$ Scuti stars exceeds this limit. All these means that the spectra observed are folded and the peaks higher than the Nyquist frequency produce false signals in the infraNyquist regime.

\paragraph{}\citet{Murphy2013} proposed a way to distinguish infra- and superNyquist modes in \textit{Kepler}'s light curves thanks to their periodically modulated sampling. Because of the spectral window that this kind of cadence produces, the superNyquist peaks split into a multiplet whose separation is a multiple of the orbital frequency of the satellite ($\approx$31nHz). For this reason, the amplitude of the peak in a non-correct regime will be lower than the real peak. We used this property to differentiate the folded peaks from the real peaks.

\begin{figure}
\resizebox{\hsize}{!}{\includegraphics{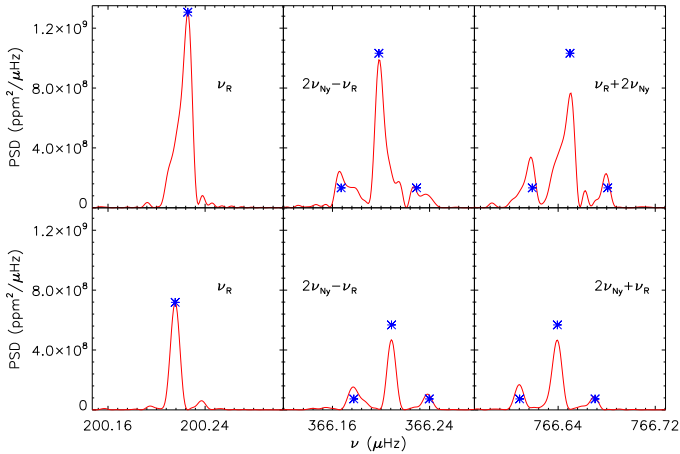}}
\caption{Upper panel from left to right: power spectral density of the $f_{1}$ mode in the infra- and two consecutive superNyquist regimes. The same for lower panel after subtracting the central peak, which allows us to observe $f_{6}$. The theoretical maximum PSD of the peaks is marked (blue asterisks) assuming that the infraNyquist peak is real (see text).}
\label{f:NRAM1}
\end{figure}

\paragraph{}Fitting the known peaks to their folded and unfolded frequencies we find their amplitudes and phases in infra- and superNyquist regime. We compare their amplitudes with a 3$\sigma$ error to ensure which peak belongs to one or the other regime. The extraction of a superNyquist peak also subtracts the other components of the multiplet. For example, we observe in Fig. \ref{f:NRAM1} that the power spectral density (PSD) of $f_{1}$ is higher in the infraNyquist regime than in the two successive superNyquist regimes. The multiplet of the peak also appears in these higher frequency regimes. The same is observed for the PSD of the highest amplitude first order peak ($f_{6}$) after subtracting the central peak (see bottom pannel). Therefore, we can conclude that these are infraNyquist peaks.

\paragraph{}According to our analysis, and considering frequencies up to $3\nu_{Ny}$ (849$\mu$Hz), we find that the three highest modes correspond to infraNyquist peaks and their modulation are not artifacts.

\subsubsection{Rotational coupling}
\label{s:rotcoupling}

\paragraph{}Another mechanism that can produce amplitude variations with time is the rotational coupling \citep{Buchler1995}. It is possible that the modes we observe are a multiplet due to rotation and that they are coupled, accomplishing the near resonance condition
\begin{equation}
2\cdot \nu_{3} +\delta = \nu_{1}+\nu_{2} \; .
\label{e:nrcond2}
\end{equation}

\paragraph{}These modes are in the resonant regime depending on their coupling coefficients \citep{Buchler1997},i.e.,
\begin{equation}
R \equiv \dfrac{|\delta|}{\gamma} \lesssim 1 - 10 \, .
\label{e:regimes}
\end{equation}

\paragraph{}The frequency mismatch for the three highest amplitude modes ($f_{1}$, $f_{2}$, and $f_{3}$) would be high ($\delta\sim7\mu$Hz). Their observed driving rates ($\gamma$) are around 8 nHz (see Table \ref{t:modpar}) and the ratio between coupling coefficients would be $R\approx10^{3}$. This value would correspond to a case far away from the resonant regime. Therefore, rotational coupling is not the cause of the observed modulation.

\section{Discussion}
\label{s:discussion}

\paragraph{}We have seen in Section \ref{s:coupling} that the individual behaviour of the parameters in each mode could be explained with resonant mode coupling. In addition, we observe that the amplitude and phase variations of the three highest modes are shifted in time. This suggests that an exchange of energy might occur between them. A direct coupling between these three modes was discarded in Section \ref{s:ocauses}, but many options are still possible. One of them implies that two unstable modes would be coupled with one stable mode. In that case, we are not able to reject a change in the behaviour of the coupling predicted by \citet{Moskalik1985}. 

\paragraph{}Another possibility that we could imagine is that these child modes exchange energy through their parent modes, possibly having some in common. The interaction between several pairs of lower frequency g modes and an acoustic mode with similar coupling coefficients has the same behaviour as a single pair \citep{Nowakowski2005}.

\begin{figure}
\resizebox{\hsize}{!}{\includegraphics{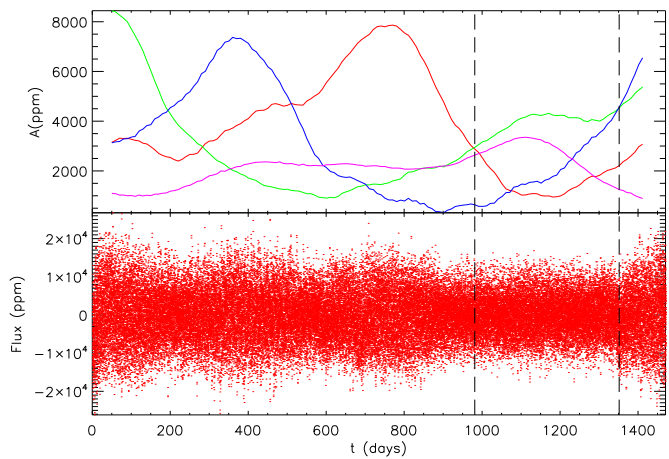}}
\caption{Top panel: amplitude variations in time. Red, green, blue, and purple indicate modes $f_{1}$, $f_{2}$, $f_{3}$, and $f_{8}$, respectively. Bottom panel: flux variations in time.}
\label{f:4th}
\end{figure}

\paragraph{}We noticed that after the successive amplitude maxima of modes $f_{1}$, $f_{2}$, and $f_{3}$, there is a time interval with no maxima and a lower dispersion in the light curve (see Fig. \ref{f:4th}). In that time interval, there is a mode, $f_{8}$, whose amplitude modulation has a maximum and follows the pattern of the shifted peaks. Nevertheless, its amplitude increase is not as much as the others. The end of the data suggest that new maxima of the three highest amplitude modes are coming.

\paragraph{}Further work can be done exploring mode coupling in more detail. Observations in other stars could guide us to determine the next step to follow.

\section{Conclusions}
\label{s:conclusion}

\paragraph{}We have observed an amplitude modulation and phase change in the three highest amplitude modes of KIC~5892969. We have proven that these variations are due to resonant mode coupling of lower frequency parent modes with one unstable child mode as suggested by \cite{Moskalik1985}. The observed characteristic parameters of this modulation are consistent with those estimated for $\delta$ Scuti stars.

\paragraph{}We see slighter amplitude and phase variations on other modes that could be interpreted as modulation due to the same mechanism with a lower ratio between the frequency mismatch and the parents driving rate \citep{Moskalik1985}. We also see modes with shorter modulation periods or more than one maximum per cycle with different height \citep{Wersinger1980}. All these modes show behaviours that are consistent with different flavours of coupling.

\paragraph{}We conclude that the resonant mode coupling seems to be the main mechanism for amplitude limitation in oscillating peaks of KIC~5892969. Furthermore, other $\delta$ Scuti stars like KIC~7106205 \citep{Bowman2014} or 4CVn \citep{Breger2000a}, show this kind of variations in one or several modes. These elements suggest that resonant coupling might be a widespread phenomena in $\delta$ Scuti stars and a major source of limitation of their mode amplitudes.

\paragraph{}
\begin{acknowledgements}
Comments from K. Belkacem, M.J. Goupil and R. Samadi are gratefully acknowledged. The authors wish to thank the \textit{Kepler} Team whose efforts made these results possible. Funding for this Discovery mission is provided by NASA's Science Mission Directorate. SBF has received financial support from the Spanish Ministry of Science and Innovation (MICINN) under the grant AYA2010-20982-C02-02. RAG acknowledge the support of the CNES \textit{CoRoT} grant.
\end{acknowledgements}

\bibliographystyle{aa}
\bibliography{citabase}

\begin{thebibliography}{31}
\expandafter\ifx\csname natexlab\endcsname\relax\def\natexlab#1{#1}\fi

\bibitem[{{Baglin} {et~al.}(2006){Baglin}, {Auvergne}, {Barge}, {Deleuil},
  {Catala}, {Michel}, {Weiss}, \& {COROT Team}}]{Baglin2006}
{Baglin}, A., {Auvergne}, M., {Barge}, P., {et~al.} 2006, in ESA Special
  Publication, Vol. 1306, ESA Special Publication, ed. M.~{Fridlund},
  A.~{Baglin}, J.~{Lochard}, \& L.~{Conroy}, 33

\bibitem[{{Balona} \& {Dziembowski}(2011)}]{Balona2011}
{Balona}, L.~A. \& {Dziembowski}, W.~A. 2011, \mnras, 417, 591

\bibitem[{{Borucki} {et~al.}(2010){Borucki}, {Koch}, {Basri}, {Batalha},
  {Brown}, {Caldwell}, {Caldwell}, {Christensen-Dalsgaard}, {Cochran},
  {DeVore}, {Dunham}, {Dupree}, {Gautier}, {Geary}, {Gilliland}, {Gould},
  {Howell}, {Jenkins}, {Kondo}, {Latham}, {Marcy}, {Meibom}, {Kjeldsen},
  {Lissauer}, {Monet}, {Morrison}, {Sasselov}, {Tarter}, {Boss}, {Brownlee},
  {Owen}, {Buzasi}, {Charbonneau}, {Doyle}, {Fortney}, {Ford}, {Holman},
  {Seager}, {Steffen}, {Welsh}, {Rowe}, {Anderson}, {Buchhave}, {Ciardi},
  {Walkowicz}, {Sherry}, {Horch}, {Isaacson}, {Everett}, {Fischer}, {Torres},
  {Johnson}, {Endl}, {MacQueen}, {Bryson}, {Dotson}, {Haas}, {Kolodziejczak},
  {Van Cleve}, {Chandrasekaran}, {Twicken}, {Quintana}, {Clarke}, {Allen},
  {Li}, {Wu}, {Tenenbaum}, {Verner}, {Bruhweiler}, {Barnes}, \&
  {Prsa}}]{Borucki2010}
{Borucki}, W.~J., {Koch}, D., {Basri}, G., {et~al.} 2010, Science, 327, 977

\bibitem[{{Bowman} \& {Kurtz}(2014)}]{Bowman2014}
{Bowman}, D.~M. \& {Kurtz}, D.~W. 2014, \mnras, 444, 1909

\bibitem[{{Breger}(1990)}]{Breger1990}
{Breger}, M. 1990, \aap, 240, 308

\bibitem[{{Breger}(2000{\natexlab{a}})}]{Breger2000}
{Breger}, M. 2000{\natexlab{a}}, in Astronomical Society of the Pacific
  Conference Series, Vol. 210, Delta Scuti and Related Stars, ed. M.~{Breger}
  \& M.~{Montgomery}, 3

\bibitem[{{Breger}(2000{\natexlab{b}})}]{Breger2000a}
{Breger}, M. 2000{\natexlab{b}}, \mnras, 313, 129

\bibitem[{{Breger} \& {Montgomery}(2000)}]{DSRS}
{Breger}, M. \& {Montgomery}, M., eds. 2000, Astronomical Society of the
  Pacific Conference Series, Vol. 210, {Delta Scuti and Related Stars}

\bibitem[{{Breger} \& {Montgomery}(2014)}]{Breger2014}
{Breger}, M. \& {Montgomery}, M.~H. 2014, \apj, 783, 89

\bibitem[{{Buchler} {et~al.}(1997){Buchler}, {Goupil}, \&
  {Hansen}}]{Buchler1997}
{Buchler}, J.~R., {Goupil}, M.-J., \& {Hansen}, C.~J. 1997, \aap, 321, 159

\bibitem[{{Buchler} {et~al.}(1995){Buchler}, {Goupil}, \&
  {Serre}}]{Buchler1995}
{Buchler}, J.~R., {Goupil}, M.~J., \& {Serre}, T. 1995, \aap, 296, 405

\bibitem[{{Dziembowski}(1982)}]{Dziembowski1982}
{Dziembowski}, W. 1982, \actaa, 32, 147

\bibitem[{{Garc{\'{\i}}a} {et~al.}(2011){Garc{\'{\i}}a}, {Hekker}, {Stello},
  {Guti{\'e}rrez-Soto}, {Handberg}, {Huber}, {Karoff}, {Uytterhoeven},
  {Appourchaux}, {Chaplin}, {Elsworth}, {Mathur}, {Ballot},
  {Christensen-Dalsgaard}, {Gilliland}, {Houdek}, {Jenkins}, {Kjeldsen},
  {McCauliff}, {Metcalfe}, {Middour}, {Molenda-Zakowicz}, {Monteiro}, {Smith},
  \& {Thompson}}]{Garcia2011}
{Garc{\'{\i}}a}, R.~A., {Hekker}, S., {Stello}, D., {et~al.} 2011, \mnras, 414,
  L6

\bibitem[{{Garc{\'{\i}}a} {et~al.}(2014){Garc{\'{\i}}a}, {Mathur}, {Pires},
  {R{\'e}gulo}, {Bellamy}, {Pall{\'e}}, {Ballot}, {Barcel{\'o} Forteza},
  {Beck}, {Bedding}, {Ceillier}, {Roca Cort{\'e}s}, {Salabert}, \&
  {Stello}}]{Garcia2014}
{Garc{\'{\i}}a}, R.~A., {Mathur}, S., {Pires}, S., {et~al.} 2014, \aap, 568,
  A10

\bibitem[{{Handler} {et~al.}(2000){Handler}, {Arentoft}, {Shobbrook}, {Wood},
  {Crause}, {Crake}, {Podmore}, {Habanyama}, {Oswalt}, {Birch}, {Lowe},
  {Sterken}, {Meintjes}, {Brink}, {Claver}, {Medupe}, {Guzik}, {Beach},
  {Martinez}, {Leibowitz}, {Ibbetson}, {Smith}, {Ashoka}, {Raj}, {Kurtz},
  {Balona}, {O'Donoghue}, {Costa}, \& {Breger}}]{Handler2000}
{Handler}, G., {Arentoft}, T., {Shobbrook}, R.~R., {et~al.} 2000, \mnras, 318,
  511

\bibitem[{{Huber} {et~al.}(2014){Huber}, {Silva Aguirre}, {Matthews},
  {Pinsonneault}, {Gaidos}, {Garc{\'{\i}}a}, {Hekker}, {Mathur}, {Mosser},
  {Torres}, {Bastien}, {Basu}, {Bedding}, {Chaplin}, {Demory}, {Fleming},
  {Guo}, {Mann}, {Rowe}, {Serenelli}, {Smith}, \& {Stello}}]{Huber2014}
{Huber}, D., {Silva Aguirre}, V., {Matthews}, J.~M., {et~al.} 2014, \apjs, 211,
  2

\bibitem[{{Mantegazza} {et~al.}(2012){Mantegazza}, {Poretti}, {Michel},
  {Rainer}, {Baudin}, {Garc{\'{\i}}a Hern{\'a}ndez}, {Semaan}, {Alvarez},
  {Amado}, {Garrido}, {Mathias}, {Moya}, {Su{\'a}rez}, {Auvergne}, {Baglin},
  {Catala}, \& {Samadi}}]{Mantegazza2012}
{Mantegazza}, L., {Poretti}, E., {Michel}, E., {et~al.} 2012, \aap, 542, A24

\bibitem[{{Moskalik}(1985)}]{Moskalik1985}
{Moskalik}, P. 1985, \actaa, 35, 229

\bibitem[{{Murphy} {et~al.}(2014){Murphy}, {Bedding}, {Shibahashi}, {Kurtz}, \&
  {Kjeldsen}}]{Murphy2014}
{Murphy}, S.~J., {Bedding}, T.~R., {Shibahashi}, H., {Kurtz}, D.~W., \&
  {Kjeldsen}, H. 2014, \mnras, 441, 2515

\bibitem[{{Murphy} {et~al.}(2013){Murphy}, {Shibahashi}, \&
  {Kurtz}}]{Murphy2013}
{Murphy}, S.~J., {Shibahashi}, H., \& {Kurtz}, D.~W. 2013, \mnras, 430, 2986

\bibitem[{{Nowakowski}(2005)}]{Nowakowski2005}
{Nowakowski}, R.~M. 2005, \actaa, 55, 1

\bibitem[{{Pamyatnykh}(1999)}]{Pamyatnykh1999}
{Pamyatnykh}, A.~A. 1999, \actaa, 49, 119

\bibitem[{{Ponman}(1981)}]{Ponman1981}
{Ponman}, T. 1981, \mnras, 196, 583

\bibitem[{{Poretti} {et~al.}(2009){Poretti}, {Michel}, {Garrido},
  {Lef{\`e}vre}, {Mantegazza}, {Rainer}, {Rodr{\'{\i}}guez}, {Uytterhoeven},
  {Amado}, {Mart{\'{\i}}n-Ruiz}, {Moya}, {Niemczura}, {Su{\'a}rez}, {Zima},
  {Baglin}, {Auvergne}, {Baudin}, {Catala}, {Samadi}, {Alvarez}, {Mathias},
  {Papar{\`o}}, {P{\'a}pics}, \& {Plachy}}]{Poretti2009}
{Poretti}, E., {Michel}, E., {Garrido}, R., {et~al.} 2009, \aap, 506, 85

\bibitem[{{Royer} {et~al.}(2007){Royer}, {Zorec}, \& {G{\'o}mez}}]{Royer2007}
{Royer}, F., {Zorec}, J., \& {G{\'o}mez}, A.~E. 2007, \aap, 463, 671

\bibitem[{{Schmid} {et~al.}(2014){Schmid}, {Theme{\ss}l}, {Breger}, {Degroote},
  {Aerts}, {Beck}, {Tkachenko}, {Van Reeth}, {Bloemen}, {Debosscher},
  {Castanheira}, {McArthur}, {P{\'a}pics}, {Fritz}, \& {Falcon}}]{Schmid2014}
{Schmid}, V.~S., {Theme{\ss}l}, N., {Breger}, M., {et~al.} 2014, \aap, 570, A33

\bibitem[{{Shibahashi} \& {Kurtz}(2012)}]{ShK2012}
{Shibahashi}, H. \& {Kurtz}, D.~W. 2012, \mnras, 422, 738

\bibitem[{{Thompson} {et~al.}(2013){Thompson}, {Christiansen}, {Jenkins}, \&
  {Haas}}]{Thompson2013}
{Thompson}, S.~E., {Christiansen}, J.~L., {Jenkins}, J.~M., \& {Haas}, M.~R.
  2013, Kepler Data Release 21 Notes

\bibitem[{{Vandakurov}(1979)}]{Vandakurov1979}
{Vandakurov}, Y.~V. 1979, \sovast, 23, 421

\bibitem[{{Wersinger} {et~al.}(1980){Wersinger}, {Finn}, \&
  {Ott}}]{Wersinger1980}
{Wersinger}, J.-M., {Finn}, J.~M., \& {Ott}, E. 1980, Physics of Fluids, 23,
  1142

\bibitem[{{Zwintz} {et~al.}(2013){Zwintz}, {Fossati}, {Guenther},
  {Ryabchikova}, {Baglin}, {Themessl}, {Barnes}, {Matthews}, {Auvergne},
  {Bohlender}, {Chaintreuil}, {Kuschnig}, {Moffat}, {Rowe}, {Rucinski},
  {Sasselov}, \& {Weiss}}]{Zwintz2013}
{Zwintz}, K., {Fossati}, L., {Guenther}, D.~B., {et~al.} 2013, \aap, 552, A68

\end{thebibliography}

\appendix
\section{Testing the methodology with an artificial $\delta$~Scuti star light curve}
\label{ap:ALC}

\paragraph{}To test the method described in Sect. \ref{s:method}, an artificial light curve is built considering Eq. \ref{e:flux}. The sampling that has been used is the same as in KIC~5892969 data in order to test if our analysis could cope with the gaps and sinusoidal variation of the Kepler's long cadence \citep{Garcia2014}. One thousand oscillations of a previous studied $\delta$ Scuti star (HD~50870) have been taken into account. The frequency range of these oscillations include peaks with typical frequencies for $\delta$ Scuti stars (from 58 to 580 $\mu$Hz) as well as lower frequencies that could be considered red noise (from 0.14 to 58 $\mu$Hz). The minimum amplitude used to built this light curve is 10 ppm and the maximum is around 12000 ppm. The phases are randomly distributed.

\begin{figure}
\resizebox{\hsize}{!}{\includegraphics{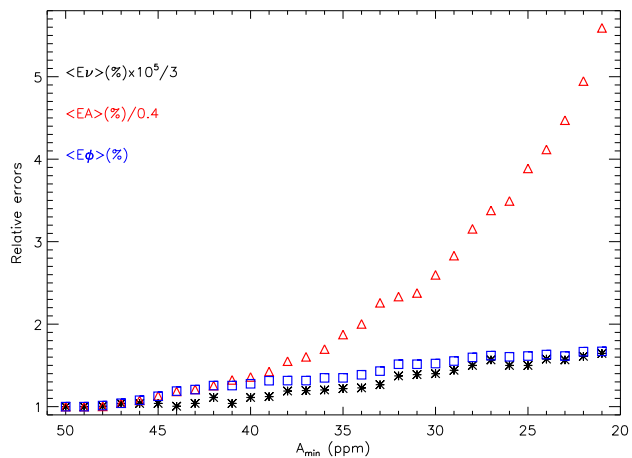}}
\caption{Evolution of the mean relative error of the frequencies (black asterisks), amplitudes (red triangles) and phases (blue squares) taking lower amplitude peaks into account (see text). Note that the curves have been conveniently scaled to the graph.}
\label{f:tEvA}
\end{figure}

\paragraph{}Since the values of the input parameters of the oscillations are known, the relative error of the outputs could be calculated. For the 170 peaks with amplitudes higher than 50 parts per million and frequencies higher than 1 $\mu$Hz, the mean relative errors obtained are 3$\times 10^{-5}$\% in frequency, 0.4\% in amplitude, and 1\% in phase. A slow increase of the mean relative error could be observed when lower amplitude peaks are taken into account (see Fig. \ref{f:tEvA}), especially for frequencies and phases. Therefore, it can be concluded that the method achieves accurate results and allows us to study real light curves of $\delta$ Scuti stars.

\end{document}